\PassOptionsToPackage{table}{xcolor}
\documentclass[sigconf]{acmart}

\setcitestyle{square,sort,comma,numbers}

\usepackage[utf8]{inputenc} %
\usepackage[T1]{fontenc}    %
\usepackage{hyperref}       %
\usepackage{url}            %
\usepackage{booktabs}       %
\usepackage{amsfonts}       %
\usepackage{nicefrac}       %
\usepackage{microtype}      %
\usepackage{xcolor} %
\usepackage[table]{xcolor}

\usepackage{amstext}
\usepackage{amsmath}
\usepackage{enumitem}
\usepackage{algorithm}
\usepackage{algpseudocode}
\usepackage{dblfloatfix}
\usepackage[show]{notes-alt}
\usepackage{multirow}
\usepackage{wrapfig}
\usepackage{tikz}
\usepackage{pgfplots}
\usepackage{subcaption}
\usepackage{pgfplotstable}
\usetikzlibrary{patterns}

\definecolor{MyGreen}{HTML}{548235}
\def\HiLi{\leavevmode\rlap{\hbox to 26em{\color{MyGreen!25}\leaders\hrule height .8\baselineskip depth .5ex\hfill}}}
\def\HiLii{\leavevmode\rlap{\hbox to 24.5em{\color{MyGreen!25}\leaders\hrule height .8\baselineskip depth .5ex\hfill}}}
\def\HiLiii{\leavevmode\rlap{\hbox to 24.5em{\color{MyGreen!25}\leaders\hrule height 1.6\baselineskip depth 3.0ex\hfill}}}

\newcommand{\idnt}{\phantom{ - }}
\newcommand{\sys}[1]{\textsc{Gar}\def\temp{#1}\ifx\temp\empty{}\else\raisebox{-.4ex}{\scriptsize #1}\fi}
\newcommand{\gar}{\textsc{Gar}}
\newcommand{\sgar}[1]{\textsc{Sgar}\def\temp{#1}\ifx\temp\empty{}\else\raisebox{-.4ex}{\scriptsize#1}\fi}

\newcommand{\sysbm}[1]{\sys{}${}_{BM25}$}
\newcommand{\quamsys}[1]{\textsc{Quam}\def\temp{#1}\ifx\temp\empty{}\else\raisebox{-.4ex}{\scriptsize #1}\fi}

\newcommand{\quam}{\textsc{Quam}}
\newcommand{\mpara}[1]{\medskip\noindent{\textbf{#1}}}
\newcommand{\laff}{\textsc{Laff}}
\newcommand{\setaff}{\textsc{SetAff}}

\newcommand{\affg}{\textit{affinity} }

\author{Mandeep Rathee}
\affiliation{%
  \institution{L3S Research Center}
  \city{Hannover}
  \country{Germany}  
}
\email{rathee@l3s.de}

\author{Sean MacAvaney}
\affiliation{%
  \institution{University of Glasgow}
  \city{Glasgow}
  \country{United Kingdom}
}
\email{sean.macavaney@glasgow.ac.uk}
\author{Avishek Anand}
\affiliation{%
    \institution{Delft University of Technology (TU Delft)}
    \city{Delft}
    \country{The Netherlands}
}
\email{avishek.anand@tudelft.nl}

\begin{document}

\title{\quam{}: Adaptive Retrieval through \underline{Qu}ery \underline{A}ffinity \underline{M}odelling}

\begin{abstract}

Building relevance models to rank documents based on user information needs is a central task in information retrieval and the NLP community.
Beyond the direct ad-hoc search setting, many knowledge-intense tasks are powered by a first-stage retrieval stage for context selection, followed by a more involved task-specific model.
However, most first-stage ranking stages are inherently limited by the recall of the initial ranking documents. Recently, adaptive re-ranking techniques have been proposed to overcome this issue by continually selecting documents from the whole corpus, rather than only considering an initial pool of documents. However, so far these approaches have been limited to heuristic design choices, particularly in terms of the criteria for document selection. In this work, we propose a unifying view of the nascent area of adaptive retrieval by proposing, \quam{}, a \textit{query-affinity model} that exploits the relevance-aware document similarity graph to improve recall, especially for low re-ranking budgets. Our extensive experimental evidence shows that our proposed approach, \quam{} improves the recall performance by up to 26\% over the standard re-ranking baselines. Further, the query affinity modelling and relevance-aware document graph modules can be injected into any adaptive retrieval approach. The experimental results show the existing adaptive retrieval approach improves recall by up to 12\%. The code of our work is available at \url{https://github.com/Mandeep-Rathee/quam}.

\end{abstract}


\maketitle

\section{Introduction}
\label{sec:intro}

Relevance modelling, which estimates whether documents satisfy an information need provided by a query, is a central task in information retrieval and NLP.
Many knowledge-intense tasks are powered by a first-stage retrieval/ranking stage for context selection, followed by a more involved task-specific model.
Traditional ranking models that rely on lexical matching (e.g., BM25) are efficient and well engineered based on decades of research but they exhibit the well-known \textit{vocabulary mismatch} problem due to the inherent under-specificity of queries.
Recent methods based on dense retrieval~\cite{karpukhin2020dense}, rely heavily on semantic similarity are slower, and rely on efficient yet lossy approximate nearest neighbor search.
In both ranking approaches, the common paradigm for ranking documents is based on the \textit{retrieve and re-rank} paradigm; where a first stage retrieval (lexical or dense) is followed by a more involved re-ranking stage facilitated by a contextual transformer model.
The primary objective of the first-stage retrieval is to maximize recall and efficiently filter out the most irrelevant documents.
However, a major limitation of this paradigm is that the recall of the final result list is, by definition, bounded by the recall of the first-stage retrieval.

To solve the bounded-recall problem, adaptive ranking techniques have been proposed that add additional opportunities to retrieve documents~\cite{macavaney2022adaptive,kulkarni2023lexically}.
The key idea of adaptive retrieval is based on modelling the similarities between documents in the corpus by constructing a \textit{corpus graph} offline.
During the re-ranking process, the neighbors of the top-scoring documents from the re-ranker are expanded using the corpus graph, allowing documents to be retrieved even if they were missed by the first-stage retriever.
Adaptive re-ranking algorithms typically either alternate between scoring results from the first-stage and the corpus graph~\cite{macavaney2022adaptive} or completely score the first-stage and then iteratively expand over the corpus graph~\cite{kulkarni2023lexically}.
Adaptive retrieval has shown to be successful with recall improvements of up to 11\% for cross-encoders~\cite{macavaney2022adaptive} and 15\% for bi-encoders~\cite{kulkarni2023lexically} when compared with existing methods and controlling for retrieval latency.

\begin{figure}
    \centering
 \begin{tikzpicture}
		\begin{axis}[
			width =  0.9\linewidth,
			height =0.6\linewidth,
			major x tick style = transparent,
			grid = major,
		    grid style = {dashed, gray!20},
			xlabel = {graph neighbours $k$},
			ylabel = {Recall@1000},
			title={},
            title style={yshift=-1.5ex}, %
            symbolic x coords={2,4,8,16,32,64,128},
            xtick={2,4,8,16,32,64,128},
            xtick distance=20,
            enlarge x limits=0.05,
            xlabel near ticks,
            ylabel near ticks,
            every axis x label/.style={at={(0.5, -0.08)},anchor=near ticklabel},
            every axis y label/.style={at={(-0.14, 0.5)},rotate=90,anchor=near ticklabel},
			]

		\addplot [color=black, style=densely dashed,line width = 1.5pt] table [x index=0, y index=1, col sep = comma] {plots/ablation-study/bm25_dl20_c1000_k_recall_c.txt}
        node[pos=0.0, anchor=west, yshift=7pt] {Re-ranking baseline -- BM25>>MonoT5};
			
	   \addplot [color=red, mark=triangle*,mark size=2.5pt, line width = 1.5pt] table [x index=0, y index=2, col sep = comma] {plots/ablation-study/bm25_dl20_c1000_k_recall_c.txt}node[pos=0.7, sloped, below] {\sys{}};

   	\addplot [color=green!60!black, mark=x, mark size=2.5pt, line width = 1.5pt] table [x index=0, y index=6, col sep = comma] {plots/ablation-study/bm25_dl20_c1000_k_recall_c.txt}
    node[pos=0.5, sloped, above] {\quamsys{}};
   
        \end{axis}
    \end{tikzpicture}    
    \caption{Recall comparison on \textbf{TREC DL20} dataset when the number of neighbors vary.} 
    \label{fig:gar_vs_quam_intro}    
    \vspace{-0.5cm}
    \Description{}
\end{figure}

\mpara{Limitations of current adaptive retrieval.} However, there are two major limitations of adaptive retrieval approaches considered.
Firstly, the quality of adaptive retrieval is based on the quality of the corpus graph, which has so far been constructed based on heuristic choices.
Specifically, current corpus graphs encode lexical or semantic similarities between documents and are agnostic to query-based relevance. 
This results in corpus graphs considering documents that have high similarity to potentially non-relevant content and might not result in surfacing relevant documents. 
Secondly, adaptive re-ranking algorithms like \gar{}~\cite{macavaney2022adaptive} do not consider the degree of similarity between documents during the expansion process. 
Consequently, \gar{} cannot differentiate between the degrees of relatedness of the documents---only which documents are most similar. In an extreme case, consider a document that isn't related to any other document in the corpus; \gar{} will waste time by scoring the nearest neighbors of this document, even though none of them are even related to the original documents.
This problem is further accentuated when denser corpus graphs or graphs with a larger number of neighbors. Figure~\ref{fig:gar_vs_quam_intro} shows this phenomenon in action, with \gar{}'s recall peaking at 32 neighbors per document, then dropping off as more are added.

\mpara{Improved Corpus Graph Construction.}
In this work, we solve both the problems mentioned before by first improving the quality of the corpus graph.
Toward this end, we train an edge-prediction model that predicts whether there should be an edge between a pair of potentially relevant documents by exploiting co-relevance information in ranking datasets. 
This learnt model as an additional output also provides an edge weight based on the \textit{learnt affinity} between documents that we refer to as the learnt-affinity scores or \laff{} scores in short.
Using learnt affinities, we are able to prune the original corpus graphs, leading to potential efficiency gains. The affinity scoring can also be leveraged for better candidate selection to improve overall recall.

\mpara{Query processing using Affinity modelling.}
Our second contribution is to propose an adaptive retrieval strategy called \quam{} (short for query affinity modelling) that judiciously chooses the neighborhood documents to re-rank by exploiting the affinity scores or edge weights. 
Unlike \gar{} which does not differentiate between neighbors of a relevant document, we propose a \textit{query-affinity model} that exploits the relevance-aware document affinity graph.

\mpara{Experimental Evaluation.}
We conduct an extensive experimental evaluation on TREC-DL '19 and '20 passage re-ranking tasks under multiple scenarios to show the efficacy and effectiveness of \quam{}.
Our results show that we can outperform the baselines resulting in clear recall improvements by up to 26\% and \gar{} by up to 12\%. 
Secondly, we show that our corpus graphs encode affinity scores that help not only \quam{} but also existing algorithms like \gar{}.
\gar{} improves by up to 9\% when using our corpus affinity graphs.
Finally, we show that \quam{} is robust to dense corpus graphs (see Figure~\ref{fig:gar_vs_quam_intro}) in that it can effectively choose between relevant and non-relevant neighbors, unlike \gar{}, which adds additional noise with increasing graph neighborhoods.

\mpara{Contributions.}

\begin{itemize}
    \item 
    We propose a novel approach to construct a corpus graph that faithfully encodes the co-relevance relations between documents called as the document affinity graph.
    
    \item 
    We provide concrete instantiation and a principled algorithm \quam{} for adaptive query processing.

    \item 
    We conduct extensive experimentation to show that we can outperform existing static and adaptive retrieval baselines.

\end{itemize}

The code of our work is available at \url{https://github.com/Mandeep-Rathee/quam}.

\begin{figure*}
    \centering
    \includegraphics[width=0.95\linewidth]{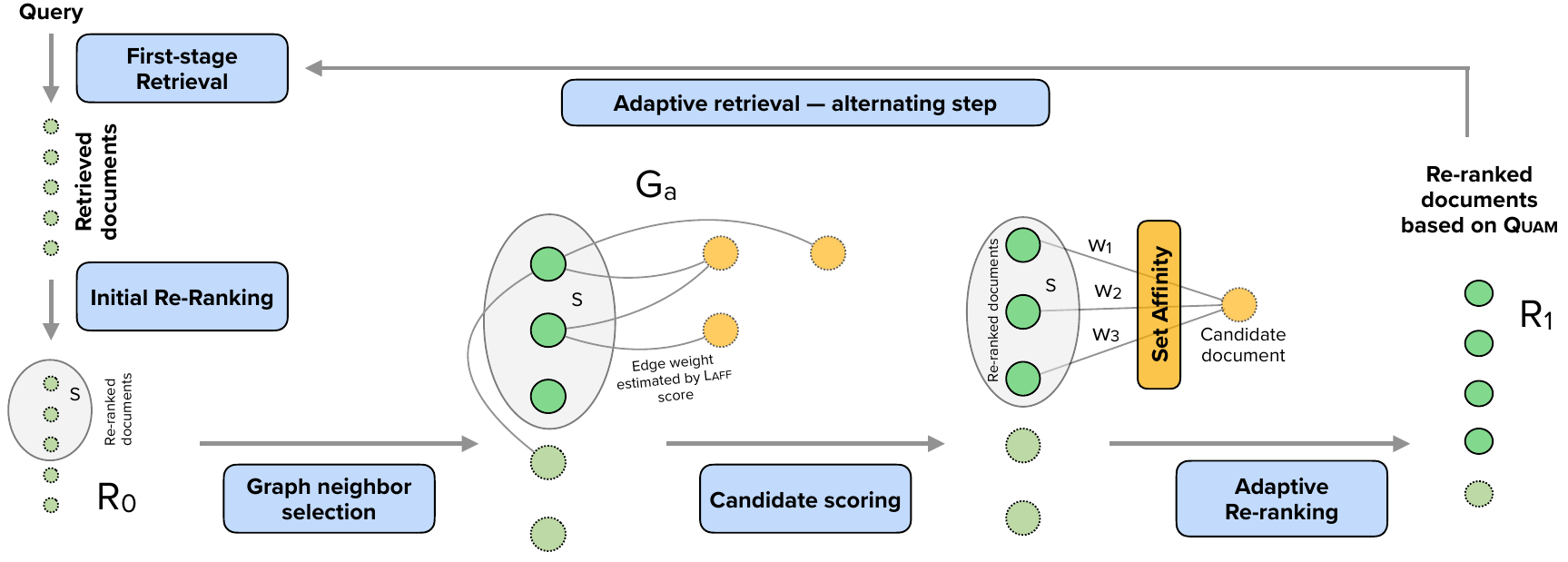}
    \caption{An overview of the adaptive retrieval through the query affinity modelling \quam{}. The $W_i$'s represent the affinity or edge weights.}
    \label{fig:laff-quam}
    \Description{}
\end{figure*}

\section{Background and Related Work}
\label{sec:relwork}

Based on the long-standing Probability Ranking Principle~\cite{robertson1977probability}, most contemporary search systems use a relevance model $\phi(q,d)$ that provides a real-valued estimate of the relevance of a document ($d$) to a query ($q$). The goal of a retrieval engine is to identify the top $k$ documents with the highest relevance score $R$ from a large corpus of documents $C$.

The trivial solution to this problem is an exhaustive search, which scores and sorts all documents in $C$. This approach is inherently unscalable since the cost increases linearly with the size of the corpus. Some types of relevance models, particularly those that use lexical~\cite{robertson2009probabilistic} (and recently learned sparse~\cite{nguyen2023unified}) representations to calculate relevance, can leverage their sparsity with inverted index data structures~\cite{zobel2006inverted} and algorithms (e.g., MaxScore~\cite{turtle1995query} and BlockMaxWAND~\cite{ding2011faster}) to reduce the cost of retrieval. These approaches are able to guarantee that the \textit{exact} top $k$ results from the corpus are returned since the relevance models behave predictably over their representations (e.g., they can often be reduced to a simple dot product between the sparse representation of the query and document).

For other types of relevance models, there are no known approaches to guarantee an exact set of top $k$ results faster than an exhaustive search. For these types of models, an \textit{approximation} of the top results is often used instead. For instance, engines that use dense bi-encoder models~\cite{karpukhin2020dense} often use algorithms such as HNSW~\cite{malkov2018efficient} to perform an approximate search. HNSW builds a neighborhood graph, where each document is linked to an approximation of its nearest neighbors. By using a hierarchy of these neighborhoods with progressively smaller subsets of the corpus, HNSW is able to scan the hierarchy to find an approximation of the most similar documents to a query vector. This technique relies on the long-standing Cluster Hypothesis~\cite{jardine1971use}, which suggests that relevant documents are likely to be near other relevant ones.

Still, approaches like HNSW are not universally applicable since they rely on scoring vectors using a simple, well-behaved function (e.g., a dot product). Many relevance models do not have this quality; for instance, learning-to-rank models often use a tree-based decision function over features~\cite{wu2010adapting}, and cross-encoder models estimate relevance through a complicated neural combination of query and document signals~\cite{nogueira2019passage}. The typical approach in this setting is to perform re-ranking~\cite{matveeva2006high} (also called cascading in literature), wherein an initial pool of $k^\prime\geq k$ documents is retrieved using a ``first-stage'' scalable approach (e.g., using the sparse or dense vector methods outlined above), and another relevance model (e.g., a learning-to-rank model or cross-encoder) then re-orders only those documents. Remarkably, re-ranking is not only helpful for learning-to-rank and cross-encoder models, but also highly competitive with HNSW for dense bi-encoder models, given the high efficiency of lexical retrieval~\cite{leonhardt2022efficient,wang2022inspection}. The key limitation of re-ranking is that it bounds performance by the documents recalled in the first stage; if a relevant document is not retrieved at that stage, it cannot be re-ranked in the final results. This limitation can be particularly problematic when the first stage uses only lexical signals, since one of the main benefits of models like cross-encoders is to overcome lexical mismatches.

Adaptive re-ranking techniques have been proposed to overcome the recall limitation of re-ranking~\cite{macavaney2022adaptive}. By using a document affinity graph and by leveraging document relevance estimations obtained during the re-ranking process, adaptive re-ranking retrieves and scores documents that were not returned during first-stage retrieval. Similar to HNSW, adaptive re-ranking leverages the Cluster Hypothesis by suggesting that documents nearby ones with high relevance estimations may also score highly. It overcomes the recall problem of re-ranking by not relying exclusively on the first stage results but instead also leveraging pseudo-relevance signals obtained during the re-ranking process itself. Several adaptive re-ranking strategies have been proposed. For instance, \gar{}~\cite{macavaney2022adaptive} alternates scoring batches between the initial pool of documents and those obtained from the document affinity graph. Follow-up work from the \gar{} authors suggests that this alternating strategy is comparable with other simple strategies for document selection from a document affinity graph~\cite{macavaney2022agent}. Beyond cross-encoders, adaptive re-ranking approaches have also been effectively applied to bi-encoders~\cite{kulkarni2023lexically,macavaney2022agent} and ensemble models~\cite{yang2024cluster}. For instance, LADR~\cite{kulkarni2023lexically} leverages an efficient lexical model to identify good ``seed'' documents to further explore its document affinity graph.

We contrast our work with prior work in two main ways. First, we replace the heuristic-based approaches of document selection, such as the alternate approach in \gar{}, with query affinity modelling to be more principled when selecting documents for scoring. Second, we replace the neighborhood graph construction process with a new learned document-document affinity model. Together, we find these changes provide more favorable selection criteria for documents and ultimately yield improved efficiency-effectiveness trade-offs.

\section{Query Affinity modelling}

In the following subsections, we focus on our proposed approach called \quam{}, specifically the two main components: 1) the document affinity graph $G_a$ based on \laff{} scores, and 2) the query affinity modelling based on \setaff{} scores. Finally, we provide a principled algorithm for using \quam{} in an adaptive retrieval pipeline. An overview of our approach is presented visually in Figure~\ref{fig:laff-quam}.
\subsection{Document Affinity Graph}
\label{sec:affinity}

Our main objective in this section is to construct document affinity graphs (in reasonable time) that help estimate relevance accurately while also improving query processing efficiency.
Using the notation presented in Section~\ref{sec:relwork}, we consider \textit{relevance} $\phi (q,d)$ a relevance model $\phi$'s estimation of the relevance of document $d$ to the information need expressed by query $q$.
In contrast, we define \textit{affinity} as the degree of association between \textit{documents} that models co-relevance. In other words, a pair of documents should have high affinity if two documents can satisfy similar information needs and low affinity if they cannot.
Existing adaptive retrieval approaches use an unweighted and undirected graph, which is called a corpus graph, denoted by $G_c$. In the case of dense retrieval, a corpus graph, a \textit{knn-graph}, can be constructed from the representation space induced by the trained document encoders. For sparse retrieval, the corpus graph can be constructed using a document as a query. The top-k-ranked result documents are its k-nearest neighbors, and the corresponding similarity scores can be used as edge weights

To construct this \affg graph $G_a$,  we start from an initial corpus graph, $G_c$, of the corpus documents and learn a model that predicts the \textit{affinity} for each of the edges in this corpus graph.
For this, we source the training data that uses \textit{query-document-label} triples to construct co-relevant document pairs that share the same query.
Consequently, we train a model \textit{f} (learnt-affinity model) that learns \textit{affinity} score between the pairs of documents. 
The affinity or edge weight between a pair of documents $d_i$ and $d_j$ in $G_{a}$ is denoted by \textit{f}($d_i$, $d_j$). Theoretically, we could use any text-matching model for $f$, though in this work, we use a cross-encoder between the two documents, given their high effectiveness at relevance modelling.

\subsection{Training Data for Learnt Affinity Model }
In some cases, large-scale datasets may already contain human-annotated co-relevance labels based on documents that are labeled as relevant to the same query. However, in practice, we posit that co-relevance labels will be sufficiently rare for training an affinity model. Indeed, Table~\ref{tab:co_relevant_pairs} shows that less than 5\% of queries in the popular MSMARCO passage train dataset~\cite{bajaj2016ms} have positive relevance labels to more than one passage, given only around 25k queries available for training. This lack of true co-relevant pairs motivates us to source pseudo co-relevant pairs. 

For each query, we source $k$ positive and negative documents. We start with a standard re-ranking approach, such as a retriever, followed by a ranker. Initially, the retriever retrieves the initial pool of documents, and then the ranker is applied to re-rank this pool. Let $R_0$ denote the pool of documents retrieved by the retriever, and $R_1$ denote the pool after it has been re-ranked by the ranker. Let $\mathcal{P}_q$ and $\mathcal{N}_q$ denote the set of relevant and non-relevant documents for the query $q$. We choose the top $k$ documents from $R_0$ as set $\mathcal{P}_q$ and the last $k$ documents as set $\mathcal{N}_q$ and the top $k$ documents from $R_1$ as set $S$. Finally, for each $p \in \mathcal{P}_{q}$, $n \in \mathcal{N}_{q}$, and $d \in S$ we have $(p, d,1) \in \mathcal{D}$ and $(n,d,0) \in \mathcal{D}$.

\begin{table}
\centering
\vspace{-1em}
\caption{Number of queries by the number of relevant documents labeled in MSMARCO passage train dataset.}\vspace{-0.75em}
\label{tab:co_relevant_pairs}
\begin{tabular}{lccccccc}
\toprule
\#rel docs & 1 & 2& 3& 4& 5& 6& 7 \\
\#query & 477580 & 21868 & 2718& 612&131 &22 & 8 \\
\bottomrule
\end{tabular}
\end{table}

We fine-tune a Bert-base\footnote{https://huggingface.co/google-bert/bert-base-uncased} model (as a cross-encoder) on the training Data $\mathcal{D}$ by minimizing the binary cross-entropy loss. 
\begin{equation}
L(\mathcal{D}) = -\frac{1}{|\mathcal{D}|} \sum_{(x,d,y) \in \mathcal{D}} \left[ y \log(f(x,d)) + (1 - y) \log(1 - f(x,d)) \right]
\end{equation}

Further details on training are available in Section~\ref{sec:aff model} of the supplementary material. Finally, we use the model $f$ to create the affinity graph $G_{a}$ by re-scoring each value in an existing corpus graph $G_{c}$.

\algdef{SE}[DOWHILE]{Do}{doWhile}{\algorithmicdo}[1]{\algorithmicwhile\ #1}%

\begin{figure}[t]\vspace{-1em}
\begin{algorithm}[H]
\caption{Adaptive Retrieval Using \quam{}} \label{alg: main}
\begin{algorithmic}
\Require Initial ranking $R_0$, batch size $b$, budget $c$, affinity graph $G_a$, top re-ranked documents $s$
\Ensure Re-Ranked pool $R_1$
\State $R_1 \gets \emptyset$ \Comment{Re-Ranking results}
\State $P \gets R_0$ \Comment{Re-ranking pool}
\State $F \gets \emptyset$ \Comment{Graph frontier}
\State $S \gets \emptyset$ \Comment{top ranked documents}
\Do
  \State $B \gets$ \Call{Score}{top $b$ from $P$, subject to $c$} \Comment{Using monoT5}
  \State $R_1 \gets R_1 \cup B$ \Comment{Add batch to results}
  \State $R_0 \gets R_0 \setminus B$ \Comment{Discard batch from initial ranking}
  \State $F \gets F \setminus B$ \Comment{Discard batch from frontier}
  \State $S \gets$ \Call{Select}{top $s$ from $R_1$} \Comment{Select top $s$ ranked docs}
  \State $F \gets F \cup (\Call{Neighbours}{B \cap S, G_a} \setminus R_1$) \Comment{Update frontier}
\State $F \gets \Call{SetAff}{d,S} \quad \forall d \in F$ \Comment{Assign set affinity scores}
  \State $P \gets \begin{cases} 
      R_0 & \text{if}\; P = F   \\
      F   & \text{if}\; P = R_0 \\
   \end{cases}$ \Comment{Alternate initial ranking and frontier}
\doWhile{$|R_1| < c$}
\end{algorithmic}
\end{algorithm} %
\end{figure}

\subsection{Query Processing Using \quam{}}
\label{sec:model}

Given a document affinity graph $G_a$, the query affinity model intends to characterize the affinity of any document to a ranked set of documents $S$. We define the expected set affinity (or \setaff{} in short) of a document $d$ from an affinity graph $G_a$ to a set of ranked documents $S$ as

\begin{equation}
\Call{SetAff}{d, S}=\sum_{d^{\prime} \in S} P(Rel(d^{\prime})) \cdot f(d,d^{\prime})
\label{eq:eaff}
\end{equation}

where $P(Rel(d^{\prime}))$ encodes the estimated relevance distribution induced by a relevance scorer (e.g., MonoT5~\cite{nogueira2020document}) model $\phi$.
$P(R(d^{\prime}))$ can be estimated in multiple ways. 
We let $P(Rel(d^{\prime})) =\frac{e^{\phi\left(q,d\right)}}{\sum_{d^{\prime} \in S} e^{\phi\left(q,d^{\prime} \right)}}$.

There are multiple methods that can be envisioned to estimate the $P(Rel(d^{\prime}))$. 
For example, one could use retrieval scores, normalized ranked positions, or re-ranking scores.
We posit that since $d^{\prime}$ is already re-ranked (i.e., $d^{\prime}\in S$), we can use the re-ranking scores as better relevance estimates in comparison to retrieval scores or rank positions. 
We test this hypothesis empirically and find that using re-ranking scores for calculating the \setaff{} yields superior performance in comparison to using retrieval scores.
We report the results for this ablation of the effect of retriever and ranker scores in Section~\ref{sec:ret vs rel} of the supplementary material.

For a given query $q$, let $R_0$ be the pool of initial ranked documents and $R_1$ be the re-ranked pool. As shown in Figure~\ref{fig:laff-quam}, the top documents from $R_0$ are used to explore the neighborhood documents in the affinity graph. These neighborhood candidate documents are assigned with set affinity scores using Equation~\ref{eq:eaff}. The high-set affinity candidate documents are selected for ranking and added to the re-ranked pool $R_1$. We keep alternating between the initial ranking and the neighbors to select the documents for re-ranking until we reach the re-ranking budget.

\subsection{Adaptive Retrieval Using \quam{}}

Alg.~\ref{alg: main} illustrates how we perform adaptive retrieval using our \quam{} model. \quam{} takes as input the initial rank pool $R_0$, a batch size $b$, a budget $c$, and an affinity graph $G_a$. Let $F$, initially empty, be the frontier that stores potential candidate documents for selection in $R_1$. Let $P$ be the re-ranking pool, initialized with $R_0$. Let $S$ be the set of top $s$ (a hyper-parameter) re-ranked documents from $R_1$. We start with selecting top $b$ (batch size) documents from $R_0$ and get relevance scores by using the \Call{Score}{ } function. These $b$ documents are added to the re-ranked pool $R_1$ and removed from $R_0$. Next, we select the set $S$ as top-$s$ re-ranked documents from $R_1$, i.e., the top $s$ documents we have re-ranked so far. Now, we use the \textit{affinity graph}, $G_a$, to explore the neighborhood documents (excluding the ones already in $R_1$) of the documents that are newly added to the set $S$. We limit the neighbor lookup to only $s$ documents because as the size of $R_1$ increases, calculating the \setaff{} scores becomes computationally expensive. The neighborhood documents are added to the frontier $F$. For each document $d$ in the frontier $F$, the set affinity score to the set $S$ (\Call{SetAff}{d, $S$}) is calculated using Equation~\ref{eq:eaff}. In contrast to \sys{}, which considers all neighbors of the source document equally important, we use these set affinity scores to prioritize the documents in the frontier $F$. 

Next, we choose the top $b$ documents from this frontier $F$. In subsequent rounds, we alternate between $R_0$ and the frontier $F$ similar to \gar~\cite{macavaney2022adaptive} until the budget criteria are fulfilled.

\section{Experimental Setup}
\label{sec:setup}
We conduct a series of experiments to answer the following research questions:
\begin{enumerate}
\item[\bf RQ1] What is the impact of \quam{} on retrieval effectiveness compared to typical re-ranking and \sys{}?
\item[\bf RQ2] How helpful is the affinity-based graph $G_a$ or \laff{} scores in prioritizing the neighbors for adaptive retrieval?
\item[\bf RQ3] What is the effect of graph depth $k$ on adaptive retrieval methods?

\item[\bf RQ4] How efficient is \quam{} in comparison to the \gar{} and standard re-ranking pipelines? 
\end{enumerate}

\subsection{Datasets and Evaluation}
We conduct our experiments mainly on the TREC Deep Learning 2019 (DL19) and 2020 (DL20) datasets~\cite{craswell2021trec} which share MSMARCO passage corpus of 8.8M passages~\cite{bajaj2016ms}. We validate our method on the DL19 and test on the DL20. The DL19 (validation) dataset consists of 43 queries with an average of 215 assessments per query. The DL20 (test) dataset consists of 54 queries with 211 relevance assessments per query. 
To evaluate the effectiveness of our approach, we use the nDCG@10, nDCG@c, and Recall@c, where $c$ is the budget. We utilize the corpus graphs generated from a sparse retriever, BM25~\cite{robertson2009probabilistic}, and a dense retriever, TCT~\cite{lin2021batch}, reusing the corpus graphs created by \sys{}.

\subsection{Ranking Models and Baselines}

For our experiments, we use different retrieval and ranking models and the most comparable adaptive retrieval baseline.

\subsubsection{Retrieval Methods} We use both sparse (BM25) and dense (TCT) retrieval models. \textbf{BM25} is a sparse retrieval method based on the query terms present in the documents. We use top $c\in[50,100,1000]$ results from BM25 using a PISA~\cite{mallia2019pisa} index. We use default parameters for retrieval. \textbf{TCT} is a dense retrieval model, a distilled version of the ColBERT model. We retrieve (exhaustively) top $c\in[50,100,1000]$ documents using the TCT-ColBERT-HNP~\cite{lin2021batch} model from \textit{huggingface}\footnote{https://huggingface.co/castorini/tct\_colbert-v2-hnp-msmarco}. 

\subsubsection{Ranking Models} We use MonoT5~\cite{nogueira2020document}\footnote{https://huggingface.co/castorini/monot5-base-msmarco} as a ranker. MonoT5 is a variant of the T5~\cite{raffel2020exploring} model, which takes a query and document as input and generates a relevance score. This score is used to re-rank the documents. In our experiments, we use the MonoT5-Base(with 223M parameters) variant trained on the MS MARCO dataset. We denote it as MonoT5 for convenience. 

\subsubsection{Baseline} We use the Graph Adaptive Retriever or \sys{}~\cite{macavaney2022adaptive} as a baseline to compare \quam{}.
\sys{} is an adaptive re-ranking approach that alternates between initial retrieved documents and neighbors of these documents in the corpus graph. Given the source document and its relevance score, \sys{} assigns the same score to all its neighbors to prioritize them.

We conduct all experiments using PyTerrier~\cite{macdonald2021pyterrier} framework. We follow the pipeline's notations from PyTerrier. For example, the pipeline "BM25>>MonoT5" retrieves using BM25 and re-ranks them using the MonoT5 model. 

\subsection{Other Hyper-parameters}
Table~\ref{tab:hyperpara} shows the hyper-parameters and their corresponding description. For Table~\ref{tab:main}, we choose batch size $b$=16, graph depth $k$=16, and vary budget $c$ from 50, 100, and 1000 and select $S$ with $s$=10, 30, and 300 respectively. We explore the robustness of our proposed method by varying batch size $b$ and graph depth $k$ in [2,128] (by power of 2). 

\begin{table}
\centering
\vspace{-1em}
\caption{Hyper-parameters and their description.}\vspace{-0.75em}
\label{tab:hyperpara}
\begin{tabular}{ll}
Notation & Description \\
\toprule
 $b$ & batch size  \\
$c$ & re-ranking budget \\
$k$ & depth of the graph (number of neighbors to explore) \\
$s$ & number of the top re-ranked documents from $R_1$\\

\bottomrule
\end{tabular}
\end{table}

\section{Results and Analysis}
\label{sec:results}

\begin{table*}
    \centering
    \caption{Effectiveness of \gar{} and \quam{}  on TREC DL19 and 20 dataset. Significant improvements (paired t-test, $p<0.05$, using Bonferroni correction) with the re-ranking baseline (retriever>>MonoT5) and \sys{} are marked with $B$ and $G$ respectively in the superscript. The best result for each pipeline is in bold.}
    {\small
    \setlength{\tabcolsep}{4.5pt}
    \begin{tabular}{llrrrrrrrrr}

\toprule
\multicolumn{1}{l}{}&\multicolumn{1}{c}{}&\multicolumn{3}{c}{$c=50$}&\multicolumn{3}{c}{$c=100$}&\multicolumn{3}{c}{$c=1000$} \\
\cmidrule(lr){3-5}\cmidrule(lr){6-8}\cmidrule(lr){9-11}
Dataset&Pipeline  & nDCG@10 & nDCG@c & Recall@c & nDCG@10 & nDCG@c  & Recall@c & nDCG@10 & nDCG@c & Recall@c\\
\midrule
\rowcolor{gray!70}
\multirow{11}{*}{\textbf{DL19}} \cellcolor{white}& MonoT5-Exh.   & 0.672 & 0.625& 0.512  & 0.672 & 0.611  & 0.599 &0.672 & 0.691 & 0.834  \\
\cmidrule{2-11}
\rowcolor{gray!50}
\cellcolor{white} &BM25>>MonoT5   &  0.676 & 0.546 &0.389 & 0.696 & 0.571 & 0.497 & 0.724  &0.696&  0.755  \\
  & \idnt w/ \sys{BM25}   &  0.694 & 0.573  & 0.426  & 0.716  & 0.605  & 0.547  & 0.719 & 0.736 & ${}^{B}$0.833\\
 & \idnt w/ \quamsys{BM25} & \bf0.706 & ${}^{BG}$\bf0.615 & ${}^{BG}$\bf0.480  & \bf0.720& ${}^{BG}$\bf0.651 & ${}^{BG}$\bf0.611  & \bf0.732 & ${}^{B}$\bf0.758  & ${}^{BG}$\bf0.867\\

\cmidrule{3-11}
 & \idnt w/ \sys{TCT}   & \bf0.724   & ${}^{B}$\bf0.620  & ${}^{B}$0.476   & ${}^{B}$\bf0.747  &  ${}^{B}$\bf0.656 & ${}^{B}$\bf0.606 &  \bf0.734  & ${}^{B}$\bf0.754 &   ${}^{B}$\bf0.859 \\
 & \idnt w/ \quamsys{TCT} & 0.704 & ${}^{B}$0.612 & ${}^{B}$\bf0.481   & 0.722& ${}^{B}$0.642 & ${}^{B}$0.601&0.721& ${}^{B}$0.747  & ${}^{B}$0.857  \\ 

\cmidrule{2-11}
\rowcolor{gray!50} 
\cellcolor{white}&  TCT>>MonoT5 & 0.732  & 0.647   & 0.506  & \bf0.722   & 0.638 &  0.610&  \bf0.699  & 0.704& 0.830  \\ 
 & \idnt w/ \sys{BM25} &  0.738  &  0.639  & 0.515    & 0.721 & 0.642  & 0.626 &  \bf0.699  & ${}^{B}$0.740 &   ${}^{B}$0.891  \\

 & \idnt w/ \quamsys{BM25} & \bf0.743& ${}^{BG}$\bf0.684 & ${}^{B}$\bf0.556  & \bf0.722& ${}^{BG}$\bf0.678 & ${}^{B}$\bf0.670 &0.696 & ${}^{B}$\bf0.741  & ${}^{B}$\bf0.896 \\

\cmidrule{3-11}
 & \idnt w/ \sys{TCT}   &  0.732  &  0.658  &  0.534 & \bf0.721  & 0.653  & 0.638 & \bf0.697  & 0.722 & 0.860 \\
 & \idnt w/ \quamsys{TCT} & \bf0.740 & ${}^{B}$\bf0.673 & \bf0.538   & \bf0.721& ${}^{B}$\bf0.667 & \bf0.659  &0.692 & ${}^{B}$\bf0.728  & \bf0.881  \\
\midrule
\rowcolor{gray!70}
\multirow{11}{*}{\textbf{DL20}} \cellcolor{white}& MonoT5-Exh.&   0.649 & 0.592& 0.576  & 0.649 & 0.593  & 0.670 & 0.649 & 0.682 & 0.852 \\
\cmidrule{2-11}
\rowcolor{gray!50}
\cellcolor{white}& BM25>>MonoT5    &  0.660 & 0.549  & 0.465  & 0.675   & 0.574 & 0.569  & \bf0.716 & 0.710 & 0.805  \\
& \idnt w/ \sys{BM25}  & 0.679  & 0.569  & 0.501 & 0.703  &  ${}^{B}$0.603  & 0.607  & 0.711   &  ${}^{B}$0.748&    ${}^{B}$0.882   \\
& \idnt w/ \quamsys{BM25}& ${}^{BG}$\bf0.716 &  ${}^{BG}$\bf0.617&  ${}^{BG}$\bf0.558  &  ${}^{B}$\bf0.715&  ${}^{BG}$\bf0.645&  ${}^{BG}$\bf0.664   & 0.707 &  ${}^{B}$\bf0.755 &  ${}^{B}$\bf0.901    \\
\cmidrule{3-11}
& \idnt w/ \sys{TCT}  &    ${}^{B}$0.720  &   ${}^{B}$0.617  &  ${}^{B}$0.570 &  \bf0.718  &  ${}^{B}$0.650  & ${}^{B}$0.688 & 0.699  & 0.740 &  ${}^{B}$0.894 \\
& \idnt w/ \quamsys{TCT}&   ${}^{B}$\bf0.727 &  ${}^{B}$\bf0.628&  ${}^{B}$\bf0.575   & 0.713&  ${}^{B}$\bf0.652&  ${}^{B}$\bf0.696  & \bf0.703 & \bf0.751 & ${}^{B}$\bf0.900  \\
\cmidrule{2-11}
\rowcolor{gray!50}
\cellcolor{white} & TCT>>MonoT5  &  0.722 &  0.642  &  0.652   & 0.701   & 0.627 & 0.713  &  \bf0.672  & 0.691&   0.848    \\
& \idnt w/ \sys{BM25}  &  \bf0.724  & 0.640  &  0.637  & \bf0.717  & 0.643  & 0.730  & 0.669  & ${}^{B}$0.720 & ${}^{B}$0.891  \\
& \idnt w/ \quamsys{BM25}  & 0.720 & ${}^{BG}$\bf0.664& \bf0.660  &0.709 & ${}^{BG}$\bf0.669 & ${}^{B}$\bf0.755  & 0.670& ${}^{BG}$\bf0.732 & ${}^{B}$\bf0.916  \\
\cmidrule{3-11}
& \idnt w/ \sys{TCT} &    \bf0.722  &  0.658  & 0.647 & 0.702  & 0.643  & 0.729   & \bf0.669  & ${}^{B}$0.707 &  0.868   \\
& \idnt w/ \quamsys{TCT}  & 0.720 & ${}^{B}$\bf0.664& \bf0.658 & \bf0.708& ${}^{BG}$\bf0.663 & ${}^{B}$\bf0.750 & \bf0.669 & ${}^{BG}$\bf0.720 & ${}^{BG}$\bf0.887    \\
\bottomrule

\end{tabular}
}
\label{tab:main}
\end{table*}

In this section, we discuss the results and analysis of our experiments. 
In all our experiments, we denote the vanilla graph-based adaptive retrieval~\cite{macavaney2022adaptive} by \sys{} and the corresponding type of corpus graph indicated in subscript, for instance, \sys{BM25} represent the graph-based adaptive retrieval method \gar{} when BM25-based corpus graph ($G_c$) is used. We denote our query affinity model as \quamsys{}, with the type of affinity graph ($G_a$) indicated in subscript. For instance, \quamsys{BM25} represents the query affinity model with the BM25-based affinity graph (i.e., the \laff{} scores from the model $f$ between pairs of documents are used to calculate the \setaff{} scores (Equation~\ref{eq:eaff})). 

\subsection{Effectiveness of \quam{}}

To answer \textbf{RQ1}, we assess the effectiveness of \quamsys{} by analyzing its impact on re-ranking pipelines with sparse (BM25) and dense (TCT) retrievals followed by a scoring function (MonoT5). We report the performance of \quamsys{} in Table~\ref{tab:main} on the TREC DL 2019 and 20. We incorporate lexical (BM25) and semantic (TCT) based corpus graphs. We compare our approach \quamsys{} with standard re-ranking pipelines and \gar{}. We vary re-ranking budgets $c$ to 50, 100, and 1000. Each row in Table~\ref{tab:main} represents a ranking system.

The standard ranking pipelines (retriever>>MonoT5) are shown in gray color. Additionally, we include the MonoT5 exhaustive (in short MonoT5-Exh.) search results for both TREC DL19 and 20. 
The recall difference between MonoT5-Exh. and the standard re-ranking pipeline BM25>>MonoT5 indicates that the BM25 retrieval fails to retrieve relevant documents that MonoT5 is capable of ranking well and hence does a poor job approximating a full MonoT5 search. This observation highlights the potential for further improvements in retrieval performance using, for instance, adaptive re-ranking techniques.

We observe the significant improvements by \quamsys{} with the affinity graph from both BM25 and TCT over the standard ranking baselines across different budget sizes. The most substantial recall improvements can be seen with a low re-ranking budget, and hence, the improved recall results in better ranking performance. In particular, in comparison to the BM25>>MonoT5 pipeline, \quamsys{BM25} improves the recall@50 from 0.389 to 0.480 (23.39\%) on DL19 and from 0.465 to 0.558 (20\%) on DL20. Similarly, \quamsys{BM25} improves the nDCG@50 by 12.64\% on DL19 and 12.39\% on DL20. \quamsys{BM25} shows similar trends as we increase the budget size. In addition, the \quamsys{TCT} improves the recall@50 from 0.465 to 0.588 (26.45\%) and recall@100 from 0.569 to 0.696 (22.32\%) on DL20. 
It is important to note that the improvements made by \sys{BM25} over the standard re-ranking baseline are not statistically significant, particularly at budget c=50 and 100. However, \quamsys{} demonstrates significant improvements over \sys{} when a sparse (BM25) retriever in combination with the BM25-based graph. In particular, \quamsys{BM25} improves recall@50 from 0.426 to 0.480 (12.68\%) on DL 19 and from 0.569 to 0.617 (11.38\%) on DL 20. For a dense (TCT) retriever, the \quamsys{BM25} significantly improves nDCG@50 and nDCG@100. 

Surprisingly, the \quamsys{BM25}outperforms the expensive MonoT5 exhaustive pipeline. In particular, \quamsys{BM25} improves recall@100 from 0.599 to 0.611 (2\%), and nDCG@100 from 0.611 to 0.651 (6.55\%), and nDCG@10 from 0.672 to 0.720 (7.14\%) on DL19. The TCT-based affinity graph also leads to similar trends. We note that MonoT5 is not an oracle relevance model; it can mistakenly assign high relevance scores to non-relevant documents. An exhaustive search setting maximizes the chances of retrieving these non-relevant documents since all documents are scored, ultimately reducing effectiveness. Meanwhile, adaptive re-ranking systems inherently constrain the search space through the initial pool and corpus graph, thereby reducing the chance of encountering this noise and resulting in higher effectiveness.

\quamsys{TCT} does not show significant improvements over \sys{TCT} (except TCT>>MonoT5 on DL20), but the performance remains comparable. Also, the \sys{TCT} does not show significant improvements over TCT>>MonoT5 pipeline, however, the \quamsys{TCT} can achieve significant improvements, especially in terms of nDCG@c. It is also important to compare both adaptive retrieval approaches with exhaustive search results. The lack of significant improvements of \quamsys{TCT} over \sys{TCT} could be due to the upper bound of the MonoT5 scoring function.

\subsection{Effect of the Affinity Graph}

To answer \textbf{RQ2}, we assess the effect of the \laff{} scores on adaptive retrieval methods. Specifically, we want to see the impact of affinity graphs $G_a$ provided to \quamsys{} and \sys{}. Towards this, we inject the BM25-based affinity graph to \sys{} which we denote by \sys{BM25}+\laff{}. An affinity graph with \sys{} adds no computational overhead since the affinity scores are pre-computed.   

We show the effectiveness of \laff{} in Figure~\ref{fig:effect of comp dl19} (on DL19) and~\ref{fig:effect of comp dl20} (on DL20). The \sys{BM25}+\laff{} shows improvements over vanilla \sys{BM25}, especially at lower re-ranking budgets. For instance, the \sys{BM25}+\laff{} improves recall@50 from 0.426 to 0.451 (5.87\%) on DL19 and 0.501 to 0.549 (9.58\%) on DL20. Similarly, it improves the nDCG@50 by 5.58\% on DL19 and 7.1\% on DL20. We observe similar trends at budget c=100.

\begin{figure}
\centering

\begin{subfigure}{0.45\columnwidth}
\begin{tikzpicture}
\begin{axis}[
    width=1.2\linewidth,
	height = \linewidth,
    ybar=0.9,
    bar width=5.0, %
    enlargelimits=0.15,
    title={},
    legend style={at={(0.5,-0.15)},
      anchor=north,legend columns=-1},
    ylabel={nDCG@c},
    ylabel near ticks,
    xlabel={budget c},
    every axis y label/.style={at={(-0.10, 0.5)},rotate=90,anchor=near ticklabel},
    symbolic x coords={50, 100, 1000},
    xtick=data,
    yticklabel style = {font=\tiny,xshift=0.5ex},
    xticklabel style = {font=\tiny,yshift=0.5ex},
    nodes near coords={},
    nodes near coords align={vertical},
    enlarge x limits=0.25,
    ]

\addplot[fill=black!60] coordinates {(50,0.546) (100,0.571) (1000,0.696)};
\addplot[pattern=north west lines, pattern color=blue!60] coordinates {(50,0.573) (100,0.605) (1000,0.736085)};
\addplot[pattern=north east lines, pattern color=red!60] coordinates {(50,0.605) (100,0.648) (1000,0.746823) };
\addplot [fill=blue!60] coordinates {(50,0.615) (100,0.651) (1000,0.758198)};

\end{axis}
\end{tikzpicture}
\end{subfigure}
\hspace{0.02\columnwidth}
\begin{subfigure}{0.45\columnwidth}
\begin{tikzpicture}
\begin{axis}[
    width=1.2\linewidth,
	height = \linewidth,
    ybar=0.9,
    bar width=5.0, %
    enlargelimits=0.15,
    title={},
    legend style={at={(0.5,-0.15)},
      anchor=north,legend columns=-1},
    ylabel={Recall@c},
    ylabel near ticks,
    xlabel={budget c},
    every axis y label/.style={at={(-0.10, 0.5)},rotate=90,anchor=near ticklabel},
    symbolic x coords={50,100, 1000},
    xtick=data,
    yticklabel style = {font=\tiny,xshift=0.5ex},
    xticklabel style = {font=\tiny,yshift=0.5ex},
    nodes near coords={},
    nodes near coords align={vertical},
    legend columns=4,
    legend entries={BM25>>MonoT5;,\sys{BM25};, \sys{BM25}+\laff;,\quamsys{BM25}},
    legend to name=compdl19,
    enlarge x limits=0.25,
    ]

\addplot[fill=black!60] coordinates {(50,0.389 ) (100,0.497) (1000,0.755495)};
\addplot[pattern=north west lines, pattern color=blue!60] coordinates {(50,0.426) (100,0.547) (1000,0.833)};
\addplot[pattern=north east lines, pattern color=red!60] coordinates {(50,0.451 ) (100,0.600) (1000,0.861)};
\addplot [fill=blue!60] coordinates {(50,0.480) (100,0.611) (1000,0.867)};

\end{axis}
\end{tikzpicture}
\end{subfigure}

\vspace{\baselineskip}
\begin{minipage}{\columnwidth}
    \begin{center}
    \ref{compdl19}
    \end{center}
\end{minipage}
\caption{Effect of Learnt Affinity (\laff) scores on adaptive retrieval on DL19 dataset.}
\label{fig:effect of comp dl19}
\Description{}
\end{figure}

\begin{figure}
\centering

\begin{subfigure}{0.45\columnwidth}
\begin{tikzpicture}
\begin{axis}[
    width=1.25\linewidth,
	height = \linewidth,
    ybar=0.9,
    bar width=5.0, %
    enlargelimits=0.15,
    title={},
    legend style={at={(0.5,-0.15)},
      anchor=north,legend columns=-1},
    ylabel={nDCG@c},
    ylabel near ticks,
    xlabel={budget c},
    every axis y label/.style={at={(-0.10, 0.5)},rotate=90,anchor=near ticklabel},
    symbolic x coords={50, 100, 1000},
    xtick=data,
    yticklabel style = {font=\tiny,xshift=0.5ex},
    xticklabel style = {font=\tiny,yshift=0.5ex},
    nodes near coords={},
    nodes near coords align={vertical},
    enlarge x limits=0.25,
    ]

\addplot[fill=black!60] coordinates {(50,0.548419 ) (100, 0.574223) (1000,0.711783)};
\addplot[pattern=north west lines, pattern color=blue!60] coordinates {(50,0.568929) (100,0.602871 ) (1000, 0.74807 )};
\addplot[pattern=north east lines, pattern color=red!60] coordinates {(50,0.602236) (100,0.64181) (1000,0.7512 ) };
\addplot [fill=blue!60] coordinates {(50, 0.616611) (100,0.645399 ) (1000,0.754889)};

\end{axis}
\end{tikzpicture}
\end{subfigure}
\hspace{0.02\columnwidth}
\begin{subfigure}{0.45\columnwidth}
\begin{tikzpicture}
\begin{axis}[
    width=1.25\linewidth,
	height = \linewidth,
    ybar=0.9,
    bar width=5.0, %
    enlargelimits=0.15,
    title={},
    legend style={at={(0.5,-0.15)},
      anchor=north,legend columns=-1},
    ylabel={Recall@c},
    ylabel near ticks,
    xlabel={budget c},
    every axis y label/.style={at={(-0.10, 0.5)},rotate=90,anchor=near ticklabel},
    symbolic x coords={50,100, 1000},
    xtick=data,
    yticklabel style = {font=\tiny,xshift=0.5ex},
    xticklabel style = {font=\tiny,yshift=0.5ex},
    nodes near coords={},
    nodes near coords align={vertical},
    legend columns=4,
    legend entries={BM25>>MonoT5;,\sys{BM25};, \sys{BM25}+\laff;,\quamsys{BM25}},
    legend to name=compdl20,
    enlarge x limits=0.25,
    ]

\addplot[fill=black!60] coordinates {(50,0.465117 ) (100, 0.56852) (1000, 0.804832)};
\addplot[pattern=north west lines, pattern color=blue!60] coordinates {(50,0.501053 ) (100,  0.606638) (1000, 0.882213 )};
\addplot[pattern=north east lines, pattern color=red!60] coordinates {(50,0.549446 ) (100,0.665206) (1000,0.895225 )};
\addplot [fill=blue!60] coordinates {(50, 0.558318 ) (100,0.663938 ) (1000,0.90053 )};

\end{axis}
\end{tikzpicture}
\end{subfigure}

\vspace{\baselineskip}
\begin{minipage}{\columnwidth}
    \begin{center}
    \ref{compdl20}
    \end{center}
\end{minipage}
\caption{Effect of Learnt Affinity (\laff) scores on adaptive retrieval on DL20 dataset.}
\label{fig:effect of comp dl20}
\Description{}
\end{figure}

\subsection{Effect of Graph Depth}
\label{sec:hyperpara}

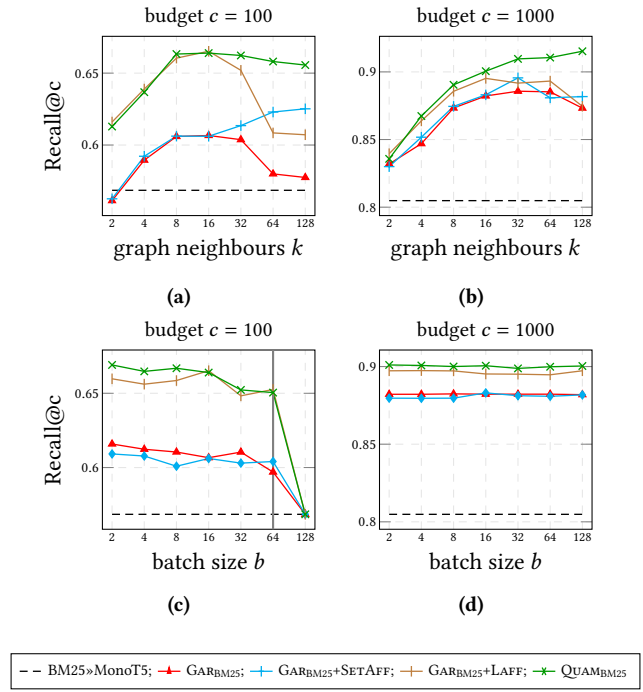
\begin{figure}
    \centering
        \begin{subfigure}{0.52\columnwidth}
        \centering
        \begin{tikzpicture}
		\begin{axis}[
			width  = \linewidth,
			height = 0.9\linewidth,
			major x tick style = transparent,
			grid = major,
		    grid style = {dashed, gray!20},
			xlabel = {graph neighbours $k$},
			ylabel = {Recall@c},
			title={\small{budget $c=100$}},
            title style={yshift=-1.5ex}, %
            legend columns=-1,
            legend entries={BM25>>MonoT5;,\sys{{\scalebox{0.7}{BM25}}};, \sys{{\scalebox{0.7}{BM25}}}+\setaff;,\sys{{\scalebox{0.7}{BM25}}}+\laff;,\quamsys{{\scalebox{0.7}{BM25}}} },
            legend style={font=\fontsize{6.0}{10}\selectfont},
            legend to name=batch_and_lk,
            legend image post style={xscale=0.5},
            symbolic x coords={2,4,8,16,32,64,128},
            xtick={2,4,8,16,32,64,128},
            xtick distance=20,
            yticklabel style = {font=\tiny,xshift=0.5ex},
            xticklabel style = {font=\tiny,yshift=0.5ex},
            enlarge x limits=0.05,
            xlabel near ticks,
            ylabel near ticks,
            every axis x label/.style={at={(0.5, -0.08)},anchor=near ticklabel},
            every axis y label/.style={at={(-0.14, 0.5)},rotate=90,anchor=near ticklabel},
			]

   			\addplot [color=black, style=densely dashed,line width = 0.5pt] table [x index=0, y index=1, col sep = comma] {plots/ablation-study/bm25_dl20_c100_k_recall_c.txt};
			
			\addplot [color=red, mark=triangle*, mark size=1.5pt,line width = 0.5pt] table [x index=0, y index=2, col sep = comma] {plots/ablation-study/bm25_dl20_c100_k_recall_c.txt};
   
     		\addplot [color=cyan, mark=+, line width = 0.5pt] table [x index=0, y index=4, col sep = comma] {plots/ablation-study/bm25_dl20_c100_k_recall_c.txt};      
      		
            \addplot [color=brown, mark=|, line width = 0.5pt] table [x index=0, y index=5, col sep = comma] {plots/ablation-study/bm25_dl20_c100_k_recall_c.txt};

   			\addplot [color=green!60!black, mark=x, line width = 0.5pt] table [x index=0, y index=6, col sep = comma] {plots/ablation-study/bm25_dl20_c100_k_recall_c.txt};

      		\end{axis}
	\end{tikzpicture}%
        \caption{}
        \end{subfigure} \hspace{-0.08\columnwidth}
        \begin{subfigure}{0.52\columnwidth}
        \centering
        \begin{tikzpicture}
		\begin{axis}[
			width  = \linewidth,
			height = 0.9\linewidth,
			major x tick style = transparent,
			grid = major,
		    grid style = {dashed, gray!20},
			xlabel = {graph neighbours $k$},
			title={\small{budget $c=1000$}},
            title style={yshift=-1.5ex}, %
            symbolic x coords={2,4,8,16,32,64,128},
            xtick={2,4,8,16,32,64,128},
            yticklabel style = {font=\tiny,xshift=0.5ex},
            xticklabel style = {font=\tiny,yshift=0.5ex},
            xlabel near ticks,
            ylabel near ticks,
            every axis x label/.style={at={(0.5, -0.08)},anchor=near ticklabel},
            every axis y label/.style={at={(-0.14, 0.5)},rotate=90,anchor=near ticklabel},
            enlarge x limits=0.05,
			]			
			\addplot [color=black, style=densely dashed,line width = 0.5pt] table [x index=0, y index=1, col sep = comma] {plots/ablation-study/bm25_dl20_c1000_k_recall_c.txt};
			
			\addplot [color=red, mark=triangle*,mark size=1.5pt, line width = 0.5pt] table [x index=0, y index=2, col sep = comma] {plots/ablation-study/bm25_dl20_c1000_k_recall_c.txt};

     		\addplot [color=cyan, mark=+, line width = 0.5pt] table [x index=0, y index=4, col sep = comma] {plots/ablation-study/bm25_dl20_c1000_k_recall_c.txt};      
      		
            \addplot [color=brown, mark=|, line width = 0.5pt] table [x index=0, y index=5, col sep = comma] {plots/ablation-study/bm25_dl20_c1000_k_recall_c.txt};

   			\addplot [color=green!60!black, mark=x, line width = 0.5pt] table [x index=0, y index=6, col sep = comma] {plots/ablation-study/bm25_dl20_c1000_k_recall_c.txt};

		\end{axis}
	\end{tikzpicture}
         \caption{}
        \end{subfigure}   
    \vfill
        \begin{subfigure}{0.52\columnwidth}
        \centering
        \begin{tikzpicture}
		\begin{axis}[
			width  = \textwidth,
			height = 0.9\textwidth,
			major x tick style = transparent,
			grid = major,
		    grid style = {dashed, gray!20},
			xlabel = {batch size $b$},
			ylabel = {Recall@c},   
			title={\small{budget $c=100$}},
            title style={yshift=-1.5ex}, %
            symbolic x coords={2,4,8,16,32,64,128},
            xtick={2,4,8,16,32,64,128},
            yticklabel style = {font=\tiny,xshift=0.5ex},
            xticklabel style = {font=\tiny,yshift=0.5ex},
            xlabel near ticks,
            ylabel near ticks,
            every axis x label/.style={at={(0.5, -0.08)},anchor=near ticklabel},
            every axis y label/.style={at={(-0.14, 0.5)},rotate=90,anchor=near ticklabel},
            enlarge x limits=0.05,
			]			
			\addplot [color=black, style=densely dashed,line width = 0.5pt] table [x index=0, y index=1, col sep = comma] {plots/ablation-study/budget-c-100/bm25_dl20_b_recall_c.txt};
			
			\addplot [color=red, mark=triangle*,mark size=1.5pt, line width = 0.5pt] table [x index=0, y index=2, col sep = comma] {plots/ablation-study/budget-c-100/bm25_dl20_b_recall_c.txt};

			\addplot [color=cyan, mark=diamond*, mark size=1.5pt,line width = 0.5pt] table [x index=0, y index=4, col sep = comma] {plots/ablation-study/budget-c-100/bm25_dl20_b_recall_c.txt};
   
      		\addplot [color=brown, mark=|, line width = 0.5pt] table [x index=0, y index=3, col sep = comma] {plots/ablation-study/budget-c-100/bm25_dl20_b_recall_c.txt};

   			\addplot [color=green!60!black, mark=x, line width = 0.5pt] table [x index=0, y index=5, col sep = comma] {plots/ablation-study/budget-c-100/bm25_dl20_b_recall_c.txt};
      
            \pgfplotsextra{
            \pgfkeysgetvalue{/pgfplots/ymin}{\ymin}
            \pgfkeysgetvalue{/pgfplots/ymax}{\ymax}
            \draw[gray, thick] (axis cs:64,\ymin) -- (axis cs:64,\ymax);
        }
      
		\end{axis}
	\end{tikzpicture}
         \caption{}
        \end{subfigure}  \hspace{-0.08\columnwidth}
    \begin{subfigure}{0.52\columnwidth}
        \centering
        \begin{tikzpicture}
		\begin{axis}[
			width  = \linewidth,
			height = 0.9\linewidth,
			major x tick style = transparent,
			grid = major,
		    grid style = {dashed, gray!20},
			xlabel = {batch size $b$},
			title={\small{budget $c=1000$}},
            title style={yshift=-1.5ex}, %
            symbolic x coords={2,4,8,16,32,64,128},
            xtick={2,4,8,16,32,64,128},
            yticklabel style = {font=\tiny,xshift=0.5ex},
            xticklabel style = {font=\tiny,yshift=0.5ex},
            xlabel near ticks,
            ylabel near ticks,
            every axis x label/.style={at={(0.5, -0.08)},anchor=near ticklabel},
            every axis y label/.style={at={(-0.14, 0.5)},rotate=90,anchor=near ticklabel},
            enlarge x limits=0.05,
			]			
			\addplot [color=black, style=densely dashed,line width = 0.5pt] table [x index=0, y index=1, col sep = comma] {plots/ablation-study/budget-c-1000/bm25_dl20_b_recall_c.txt};
			
			\addplot [color=red, mark=triangle*,mark size=1.5pt, line width = 0.5pt] table [x index=0, y index=2, col sep = comma] {plots/ablation-study/budget-c-1000/bm25_dl20_b_recall_c.txt};
	
			\addplot [color=cyan, mark=diamond*,mark size=1.5pt, line width = 0.5pt] table [x index=0, y index=4, col sep = comma] {plots/ablation-study/budget-c-1000/bm25_dl20_b_recall_c.txt};

         	\addplot [color=brown, mark=|, line width = 0.5pt] table [x index=0, y index=3, col sep = comma] {plots/ablation-study/budget-c-1000/bm25_dl20_b_recall_c.txt};

   			\addplot [color=green!60!black, mark=x, line width = 0.5pt] table [x index=0, y index=5, col sep = comma] {plots/ablation-study/budget-c-1000/bm25_dl20_b_recall_c.txt};
      
		\end{axis}
	\end{tikzpicture}
         \caption{}
        \end{subfigure}

\vspace{\baselineskip}
\begin{minipage}{\columnwidth}
\begin{center}
\ref{batch_and_lk}
\end{center}
\end{minipage}
 \normalsize
    \caption{Recall comparison on \textbf{TREC DL20} dataset when the number of neighbours $k$ (with fixed $b=16$) and batch size $b$ (with fixed $k=16$) vary. The vertical line at $b=64$ separates the region where $b>c$. } 
    \label{fig:ablation-parameter}    
  \Description{}
\end{figure}

Like in standard retrieval settings, where recall improvements can be achieved by processing higher retrieval depths, in adaptive retrieval, higher recall can be achieved by processing deeper graph neighborhoods.
However, traversing more documents either by accessing higher retrieval depths or graph neighborhoods adds more non-relevant documents that need to be differentiated from the relevant documents.
In this experiment, we closely examine the ability of adaptive retrieval methods to achieve higher recall by traversing deeper graph neighborhoods. 
Towards this, in addition to \gar{}, we re-inforce \sys{} with components of our \quam{} model -- the \laff{} scores and dynamic set affinity computation--to construct even stronger baselines. 
In Figure~\ref{fig:ablation-parameter}, we show the effect of graph depth $k$ (in the first row) to show the performance of \gar{}, \quam{} and the two baselines \gar{}+\setaff{} and \gar{}+\laff{}.

We firstly observe that there is a noticeable difference in performance between \gar{} and \quam{} at all graph depths and is magnified at higher graph depths.
In fact, the performance of \gar{} degrades substantially at higher graph depths due to its inability to differentiate between relevant or non-relevant documents.
It is also clear that \laff{} scores have a positive impact on \gar{} (similar to our observation in the last section). 
However, even \gar{}+\laff{} degrades in performance at higher graph depths.
This is mainly due to the fact that \gar{} cannot differentiate between two neighbors of the same relevant document.
Secondly, \gar{} also processes all neighbors of a re-ranked document before going to the next relevant document, introducing the risk of adding potentially non-relevant documents if the ranked document is also non-relevant.
Finally, we can also see that when \gar{} is re-inforced with the careful \setaff{} selection in the \gar{}+\setaff{} baseline, it is able to source from more relevant neighbors.
However, the inability of \gar{} to differentiate between two neighbors of the same relevant document means it still underperforms \quam{} at higher retrieval depths.

We also show the effect of the graph depth on ranking performances across different budgets in Figure~\ref{fig:ablation-dl-19} (on DL19) and~\ref{fig:ablation-dl-20} (on DL20) in Section~\ref{sec:effect_of_k_appen} of supplementary material.  

Interestingly, all approaches show insensitivity to variations in batch size. This characteristic of batch size insensitivity is advantageous as it enables the utilization of the full computational capacity of the hardware, consequently reducing latency. We show the effect of batch size on the ranking performances across different budgets in Figure~\ref{fig:ablation-batch-dl-19} (on DL19) and~\ref{fig:ablation-batch-dl-20} (on DL20) in Section~\ref{sec:effect_of_b_appen} of supplementary material.

\subsection{Efficiency of Query Processing}
\label{sec:efficiency}

We re-iterate that re-ranking pipelines using adaptive retrieval have the same number of re-ranking operations as classical re-ranking pipelines, and this cost dominates the total computational cost of the pipeline. Indeed, adaptive retrieval procedures like \gar{} and \quam{} are designed to contribute minimally to the total computational cost.
To verify this property empirically, we performed latency experiments to assess the computational overhead introduced by \quamsys{} in comparison to \gar{}. 
Note that while \gar{} indiscriminately schedules candidate documents for re-ranking, \quam{} selects documents by computing a set-affinity score for each candidate document.

For a fair comparison between \gar{} and \quam{}, we use the same MonoT5 re-ranker, a BM25-based graph of depth $k=16$ and a batch $b=16$ on the same hardware. While the MonoT5 scoring process leverages a GPU for hardware acceleration, both \gar{} and \quam{} utilize only a single thread on CPU.
In Table~\ref{tab:latency}, we report the recall and mean latency (in ms) per query at different re-ranking budgets. 
For stable measurements, we take the average over $5$ consecutive runs. 
We find that the variance in runtimes is as low as 0.01 ms, and hence we discard it in the table.

We first observe that both adaptive retrieval-based approaches, \gar{} and \quamsys{} account for only $2-3$\% of the total time taken by ranker (MonoT5). For instance, the MonoT5 takes an average of approximately $3479.66$ ms (with a batch size of $64$) to re-rank $1000$ documents per query on our hardware. On the other hand, the adaptive retrieval components of \gar{} and \quamsys{} (with batch size $b$=16) take around 97.05 and 95.72 ms, respectively. 

At lower re-ranking budgets ($c=50$ and $c=100$), the \quamsys{} takes slightly longer (an additional time of 0.74 and 1.97 ms, but with recall improvement of 12.7\% and 11.7\% respectively) to process the neighborhood documents. 
Since the \gar{} looks for neighbors of $b$ documents at each iteration, on the other hand, \quamsys{} looks for neighbors of $|S|=s$ documents and uses the Equation~\ref{eq:eaff} to compute the \setaff{} scores. 
However, as the budget $c$ increases, \quamsys{} also outperforms \gar{} in terms of speed since the number of lookups for \quamsys{} is less than that of \gar{}. The most important observation is that the \quamsys{} achieves the recall of $0.849$ in $57.36$ ms/query whereas \gar{} takes $97.05$ ms/query to obtain a recall of $0.833$. This demonstrates that, for a given sufficient budget size, \quamsys{} outperforms \gar{} in both the quality and latency of adaptive retrieval. 
In conclusion, we see that the computational overheads for \quam{} is comparable or sometimes better than \gar{} while delivering consistently better recall at all re-ranking budgets.

\newcommand{\gt}[1]{\tiny{(\tikz\draw[green!60!black, fill=green!60!black] (0,0) -- (0.10,0) -- (0.05,0.10) -- cycle;{#1\%}})}

\newcommand{\rt}[1]{\tiny{(
    \tikz\draw[red, fill=red] (0,0) -- (0.10,0) -- (0.05,-0.10) -- cycle;%
    {#1\%})}}

\begin{table} %
\centering
\caption{Mean latency overheads for \quamsys{} and \sys{}  (ms/query). $|S|$ denotes the size of set $S$ in Equation~\ref{eq:eaff}. We denote the gain/drop in performance by \quamsys{} by a green/red triangle over the \gar{}.}\vspace{-0.5em}
\label{tab:latency}
\begin{tabular}{rrclcl}
\toprule
& &\multicolumn{2}{c}{time (ms)}& \multicolumn{2}{c}{Recall@c}  \\
\cmidrule(lr){3-4} \cmidrule(lr){5-6} 
c & $|S|$  & \sys{BM25} & \quamsys{BM25} &\sys{BM25}  & \quamsys{BM25} \\
\midrule
50    & 10  & 2.64   &3.38\rt{28}  & 0.426 & 0.480\gt{12.7} \\
100   &  30 &  5.54     & 7.51\rt{35.6}  & 0.547 &0.611\gt{11.7}  \\
250   & 50  & 19.55    & 16.78\gt{14.2}  & 0.693 & 0.742\gt{7.1} \\
500    & 100  & 44.45   & 36.06\gt{18.9}  & 0.772 & 0.821\gt{6.3}\\
750  &  150 &  69.77   &57.36\gt{17.8}  & 0.811 & 0.849\gt{4.7}\\
1000  & 300  & 97.05   &95.72\gt{1.4}  & 0.833 &0.867\gt{4.1} \\
\bottomrule
\end{tabular}\vspace{-1em}
\end{table}

\section{Conclusion}
\label{sec:Conclusion}

In this paper, we advance the new area of adaptive retrieval as a recall-improving approach for ad-hoc retrieval.
We improve on the heuristic graph construction approaches used in earlier works by constructing an affinity graph with learned edge weights based on co-relevance information.
Additionally, we propose a new adaptive retrieval method or \quam{} that effectively chooses potential relevant documents from the affinity graph.
Our experiments clearly show that our affinity modeling for graph construction and query processing improves not only our proposed approaches but also existing adaptive retrieval approaches.
Secondly, we show that \quam{} is able to judiciously filter out non-relevant documents, resulting in higher recall at deeper graph neighborhoods.
Finally, we show that \quam{} has a low computational overhead in comparison to \gar{} and can be used in many low-latency use cases.
In the future, we would want to explore if one can further optimize the choice of candidate documents at low retrieval depths. 
Additionally, it would be important to extend adaptive retrieval from ad-hoc retrieval to more structured retrieval or learning to rank settings.

\bibliographystyle{ACM-Reference-Format}

\bibliography{references}
\balance

\clearpage
\appendix
\section*{Supplementary Material}

\section{Details on Dataset and Training Learnt Affinity Model}
\label{sec:aff model}
We randomly choose 50K queries from the MSMARCO train data and use TCT>>MonooT5 re-ranking pipeline where TCT~\cite{lin2021batch} is a strong dense retriever, and MonoT5~\cite{nogueira2020document} is an effective ranker. We take the top 5 documents from the retriever as positives ($\mathcal{P}_q$), the last 5 as negatives ($\mathcal{N}_q$), and take the top 5 documents from the re-ranked documents using MonoT5 as set $S$. We release the training dataset with our code.

We conducted our experiments on NVIDIA A100 GPU with 40G of RAM using the PyTorch library. We fine-tuned a bert-base model followed by a linear classifier on our training dataset. We used binary cross-entropy to train our model. We used Adam optimizer with a batch size of 8 and a learning rate 3e-7. We trained our model for 5 epochs. We also used a linear scheduler for the learning rate with warm-up steps.  We have released our model's weights along with our code for reproducibility. The affinity model $f$ predicts the similarity between a pair of documents.

\section{Effect of Relevance and Retrieval Scores}
\label{sec:ret vs rel}
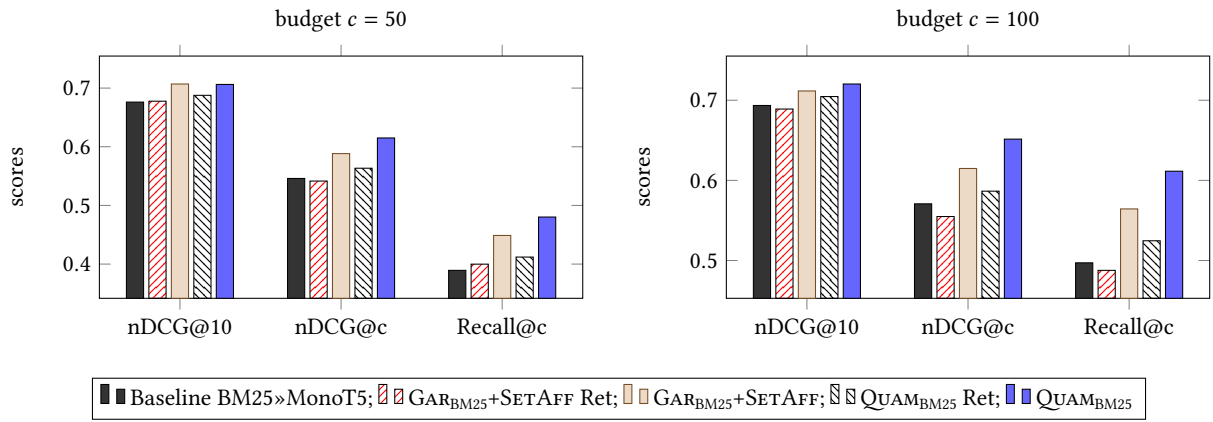
\begin{figure*}[htbp!]
\begin{subfigure}{0.45\textwidth}
\begin{tikzpicture}
\begin{axis}[
    width=\linewidth,
	height = 0.6\linewidth,
    ybar,
    bar width=6.5, %
    enlargelimits=0.15,
    title={budget $c=50$},
    legend style={at={(0.5,-0.15)},
      anchor=north,legend columns=-1},
    ylabel={scores},
    ylabel near ticks,
    every axis y label/.style={at={(-0.14, 0.5)},rotate=90,anchor=near ticklabel},
    symbolic x coords={nDCG@10,nDCG@c,Recall@c},
    xtick=data,
    nodes near coords={},
    nodes near coords align={vertical},
    enlarge x limits=0.25,
    ]
\addplot[fill=black!80] coordinates {(nDCG@10,0.67615) (nDCG@c,0.545953) (Recall@c,0.389309)};
 \addplot[pattern=north east lines, pattern color=red] coordinates {(nDCG@10,0.677649) (nDCG@c,0.541512) (Recall@c,0.399878)}; %
  \addplot coordinates {(nDCG@10,0.706954 ) (nDCG@c,0.588322) (Recall@c,0.448795)}; %
\addplot[pattern=north west lines] coordinates {(nDCG@10,0.68767) (nDCG@c,0.563403) (Recall@c, 0.411962)}; %
  \addplot[fill=blue!60] coordinates {(nDCG@10,0.70622) (nDCG@c,  0.61499) (Recall@c, 0.480199)}; %
\end{axis}
\end{tikzpicture}
\end{subfigure}
\hspace{0.02\columnwidth}
\begin{subfigure}{0.45\textwidth}
\begin{tikzpicture}
\begin{axis}[
    width=\linewidth,
	height = 0.6\linewidth,
    ybar,
    bar width=6.5, %
    enlargelimits=0.15,
    title={budget $c=100$},
    legend style={at={(0.5,-0.15)},
      anchor=north,legend columns=-1},
    ylabel={scores},
    ylabel near ticks,
    every axis y label/.style={at={(-0.14, 0.5)},rotate=90,anchor=near ticklabel},
    symbolic x coords={nDCG@10,nDCG@c,Recall@c},
    xtick=data,
    nodes near coords={},
    nodes near coords align={vertical},
    legend entries={{Baseline BM25>>MonoT5;, \sys{BM25}+\setaff{} Ret;, \sys{BM25}+\setaff;,\quamsys{BM25} Ret;, \quamsys{BM25}}},
    legend to name=retvsrel,
    enlarge x limits=0.25,
    ]
\addplot[fill=black!80] coordinates {(nDCG@10,0.693453 ) (nDCG@c,0.570714) (Recall@c,0.497101)};
\addplot[pattern=north east lines, pattern color=red] coordinates {(nDCG@10,0.689007) (nDCG@c,0.554894) (Recall@c, 0.487719)}; %
\addplot coordinates {(nDCG@10, 0.71156 ) (nDCG@c,0.614846) (Recall@c,0.564378)}; %
\addplot[pattern=north west lines] coordinates {(nDCG@10,0.704586) (nDCG@c,0.586541) (Recall@c, 0.524637)}; %
\addplot[fill=blue!60] coordinates {(nDCG@10,0.72023) (nDCG@c,0.651476) (Recall@c,0.611429)}; %
\end{axis}
\end{tikzpicture}
\end{subfigure}

\vspace{\baselineskip}
\begin{minipage}{\textwidth}
    \begin{center}
    \ref{retvsrel}
    \end{center}
\end{minipage}
\caption{Effect of using retrieval and relevance scores in \setaff{} scores on TREC DL19 dataset. Here \quamsys{BM25}-Ret and \quamsys{BM25} represent the Query Affinity Model when retrieval and relevance scores are used to calculate the \setaff{} scores respectively.}
\label{fig:ret_vs_rel}
\Description{}
\end{figure*}

According to the definition of the \setaff{} in Equation~\ref{eq:eaff}, the set affinity depends on how the relevance scorer $\phi$ estimates the query. We study the behavior of adaptive retrieval approaches under different query estimators. Toward this, we use a retriever (BM25) and a ranker (MonoT5) as query estimators. In Figure~\ref{fig:ret_vs_rel}, we report the performances of \gar{} and \quam{} with the combination of the estimator as a retriever (denoted with Ret) and a ranker. We observe that both \gar{} Ret and \quam{} Ret degrade when the retriever scores are used for the estimated relevance. This shows the retriever scores are not clean estimates of the query.  Based on empirical study, we propose to use the ranker as a true estimator of the query.  

\section{Effect of Graph Depth}
\label{sec:effect_of_k_appen}

We study the performance of different adaptive retrieval approaches when we vary the graph depth $k$ and the batch size $b$ is fixed to 16. Figure~\ref{fig:ablation-dl-19} and~\ref{fig:ablation-dl-20} show the performance comparisons for DL19 and 20 respectively.
\input{plots/ablation-dl-19}
\input{plots/ablation-dl-20}

\section{Effect of batch size}
\label{sec:effect_of_b_appen}
We also study the performance of different adaptive retrieval approaches when the batch size $b$ varies and the graph depth $k$ (=16) is fixed. Figure~\ref{fig:ablation-batch-dl-19} and~\ref{fig:ablation-batch-dl-20} show the performance comparisons for DL19 and 20 respectively.
\include{plots/ablation-batch-dl-19}
\include{plots/ablation-batch-dl-20}

\end{document}